\documentstyle[11pt,fleqn,feynman]{article}
\newcommand{\sect}[1]{ \section{#1} \setcounter{equation}{0} }

\newcommand{\beq}{\begin{equation}}
\newcommand{\eeq}{\end{equation}}
\newcommand{\beqar}{\begin{eqnarray}}
\newcommand{\eeqar}{\end{eqnarray}}

\newcommand{\eps}{\epsilon}

\newcommand{\al}{\alpha}
\newcommand{\bet}{\beta}
\newcommand{\df}{\partial}

\newcommand{\sst}{\textstyle}
\newcommand{\wh}{\widehat}

\newcommand{\Ne}{$ 1/N $~-~expansion}
\newcommand{\On}{$ O(N) $ }
\newcommand{\NN}{$ 1/N^{2} $ }
\newcommand{\N}{$ 1/N $ }
\newcommand{\AB}{$ \phi^{A}\phi^{B} $ }

\newcommand{\MS}{{\bf MS}}
\newcommand{\RG}{{\bf RG}}
\begin{document}
\hfill {\makebox[5cm]{\large \bf SPbU--IP--97--17\hfill }}
\vskip 0.2cm

\hfill {\makebox[5cm]{\large \bf NORDITA--97--69\hfill }}
\vskip 0.2cm
\hfill {\makebox[5cm]{\large  October 3, 1997\hfill }}

\vskip 2cm

\begin{center}
{\huge The simple scheme for the calculation of the anomalous
dimensions
of composite operators in  the
$ 1/N $ expansion.} \\ [8mm]
{\large S.\'{E}. Derkachov}$^a$\footnote{e-mail:
derk@nordita.dk}\footnote{On leave of absence from
St Petersburg Technology Institute,
Department of Mathematics,
Sankt Petersburg, Russia.}
{\large
\& A.N. Manashov}$^b$\footnote{e-mail: manashov@snoopy.phys.spbu.ru}
\\ [8mm]
\begin{itemize}
\item[$^a$]
{\it NORDITA, Blegdamsvej 17, DK-2100, Copenhagen, Denmark.}
\item[$^b$] {\it Department of Theoretical Physics, State University of
 St. Petersburg, \\ Sankt Petersburg, 198904 Russia.}
\end{itemize}
\end{center}

\vspace{3cm}
\begin{abstract}
The simple method for the calculating of the anomalous dimensions of
the
composite operators up to \NN\  order  is developed.
We demonstrate the effectiveness of this approach by computing the
critical exponents of the
$ (\otimes \vec \Phi)^{s}$ and
$ (\vec \Phi\otimes(\otimes \vec\df)^{n} \vec \Phi)$
operators  in the \NN\ order in the nonlinear sigma model.
The special simplifications due to the conformal invariance of the model
are discussed.
\end{abstract}

\newpage

\renewcommand{\theequation}{\thesection.\arabic{equation}}
\setcounter{equation}{0}
\section{Introduction}
The large $N$ expansion provides  the  powerful  tool for the
investigating  of the fields  theoretic models with a internal
symmetry. Being nonperturbative in its nature this method has proved to
be successful in revealing properties  which are inaccessible
in conventional perturbation theory. The \N approach relies on the
expansion of the effective action in the saddle point approximation
which determines the leading order structure in
$ 1/N$. However, beyond the
lowest order the computations become quite untractable because the
propagators of auxiliary fields depend on the mass in a very
complicated fashion.

To obviate these difficulties it had been suggested in
Refs.~\cite{V1}, where the nonlinear sigma model had been analyzed,
to consider theory directly at critical point, that should ensure the
masslessness of all propagators. The latter circumstance simplifies
considerably the computation of the Feynman diagrams. Solving the
skeleton Dyson equation the authors of Refs.~\cite{V1,V2} have
calculated the basic indices of sigma model with \NN accuracy in the
arbitrary space dimensions. Whilst the initial application of  self
consistency method was to a bosonic model, the techniques have also
been extended to examine the fermionic theories~\cite{jagF}.  The
further progress in the calculation of the critical indices is related
to the conformal bootstrap method. In the frame of this approach  the
most strong at the present moment results~---~
$ 1/N^{3} $ anomalous dimensions of the basic fields in the
$ \sigma $  model~\cite{V3} and in the Gross~-Neveu
model~\cite{V4,jagN}~---~had been obtained.

If one is interested in the calculation  of the critical indices
related to the basic fields only, these methods~--~self~-~consistency
equations and the conformal bootstrap method~--~are likely to be the
most efficient. However, some specific
features of the above methods make them inconvenient for the
treatment of the composite operators, except for the simplest
ones~\cite{V5,BGK}.  At the same time, the calculations of this type
became actual now. First of all,  it is related with the attempts to
understand in more details the structure of the conformal fields
theories (CFT) in
$ D>2 $ dimensions~(see Refs.~\cite{Osborn,Petkou,Ruhl}
and references therein).
The basic examples of
higher dimensional CFT being bosonic and fermionic
$ N $ vector models, the \N expansion underlies the computational
method used for analyzing of these models. Second, the renewing
of the interest to the \N  calculations is conditioned  by the recent
progress in the higher order computations of the renormalization group
(\RG)
functions in QCD.  The analytical calculations in QCD in the
high orders (four loop
$ \beta $~-~function~\cite{Larinbeta4},
anomalous dimensions of twist~-~2
operators~\cite{Larintwist2,Larintwist22}) are very complex ones
because of the huge number of Feynman diagrams  to be evaluated.  Thus
it is important to have methods which may ensure the independent checks
of this results.  The one of the possible approach of this type is the
$ 1/N_{f} $ expansion~\cite{jagNf}.  Further, the necessity of the
careful analysis of the anomalous dimensions of the composite operators
of special type have arisen in connection with the so~-~called
stability problem~\cite{Lerner,Wegner,Hikami}.  Recently, it was
realized  that the $ 1/N $ expansion is the more suitable approach for
this purpose~\cite{SS}.

As mentioned above, the  calculations in
\N scheme beyond the \N order become tractable only when theory is
considered directly at the critical point. But in this case the
corresponding model lost the property of the multiplicative
renormalizability and cannot be worked out with the help of the
standard \RG\ methods. It forces one to use other approaches applicable
in this case. The  methods of self~-~consistency
equations and conformal bootstrap, discussed above, are applicable for
the operators of the special type only. The approach based on the
operator product expansion used in Refs~\cite{Ruhl,Petkou} is too
involved to have a practical use in higher order calculations and
limited by the models possessing of the conformal symmetry.
The most convenient approach for the calculating of the anomalous
dimensions of the composite operators in the first order of \N
expansion had been developed in Ref~\cite{Step}. The authors
of Ref.~\cite{Step} have exploited the
property of the scale  invariance  of the correlators at critical point
to obtain the anomalous dimensions of arbitrary composite operators.
The main result of the paper~\cite{Step} is the formula expressing
\N  order anomalous dimensions
via the
$ \Delta $ pole residues of the corresponding
diagrams~($ \Delta $ is the regularization parameter, analogous
$ \epsilon $ in dimensional regularization). Certainly, this way is the
most effective in the case of the operators of arbitrary type.

At the straightforward generalization beyond the \N order
(see Refs.~\cite{GraceyA,SS}) the above method lost many its attractive
features. In the present paper we shall take advantage of the other
approach~\cite{Nal} which allow to derive the simple formula for the
anomalous dimensions in the \NN order.
To avoid the
inconveniences connected with the absence of the multiplicative
renormalizability we consider  extended model which is reduced to the
initial one by tuning of some parameters. The extended model can be
analyzed by the standard renormalization group (\RG) methods.
To obtain  the anomalous dimensions in the
initial model we relate it  with the corresponding
\RG~-~functions  of the extended one.  In the following we restrict
ourselves to the nonlinear sigma model, the generalization on the other
model being straightforward.

The paper is organized  as follows: In section 2 we review the basic
features of the nonlinear sigma model and discuss the problem to be
solved. In section 3 the basic formulae for the anomalous dimensions
of composite operators are derived. Some technical tricks useful in
higher order calculations are collected in section 4. The results of
\NN order calculations of anomalous dimensions of some composite
operators are presented in  section 5. The conclusions are given in
section 6.

\setcounter{equation}{0}

\section{Preliminaries}
\label{prel}
The \Ne\ for the massless \On--nonlinear sigma model in
$D\equiv 2\mu$-dimensions is derived in the standard manner from
the following action:
\begin{equation}
S = -\frac{1}{2}(\df\phi^{A})^{2}
 -\frac{1}{2}M^{-2\Delta}\psi K_{\Delta} \psi
 +\frac{1}{2}\psi(\phi^{A})^{2}
 +\frac{1}{2}\psi K \psi,
\label{action}
\end{equation}
where
$ \phi^{A} $  is the \On vector field and
$ \psi $--auxiliary field;
$\Delta$ - regularization parameter,
$ M $ renormalization mass.
The two first term in~(\ref{action}) form the free part of the action,
whereas two last are treated as a interaction.
The kernel $K$ is determined from the requirement of the
cancellation of
the simple
$ \phi $ loop insertion in
$ \psi $ line
\begin{center}
\Lengthunit=0.7cm
\GRAPH(hsize=10){\mov(6,0){\dashlin(2,0)\mov(1,0){\Circle**(0.3)}
\mov(2,-0.15){$+ \  \frac{\textstyle N }{\textstyle 2}$}
\mov(7,-0.15){$\textstyle = 0$}
\mov(4.5,0){\mov(0.95,0){\dashlin(1,0)}
\mov(-0.75,0){\dashlin(-1,0)}\Circle(1.25)}}}
\vspace{0.3cm}
\parbox[t]{140mm}
{}
\end{center}
\vskip -0.5cm
and reads as:
\begin{equation}
K(p) = -\frac{N}{2}
\int \frac{{\rm d}^{D}k}{{(2\pi)}^{D}}
\frac{1}{k^2(p-k)^2}
=-\frac{N}{2(4\pi)^{\mu}}\frac{H^{2}(1)}{H(2\mu-2)} p^{D-4}
\equiv b^{-1}p^{D-4},
\label{Kp}
\end{equation}
where $H(x)\equiv\Gamma(\mu-x)/\Gamma(x)$.

The regularized kernel
$ K_{\Delta} $ is defined as
$
K_{\Delta} = C^{-1}(\Delta)K(p)p^{2\Delta}.
$
The function
$ C(\Delta) $  is assumed to be regular at
$ \Delta=0 $ and
$ C(\Delta)\to 1 $ at
$ \Delta\to 0$. The freedom in the choice of the form of
$ C(\Delta) $ will be discussed in Sec.{\ref{extend}}.
The propagators of $\phi$ and
$ \psi $ fields have form
~($G_{\psi}=K^{-1}_{\Delta}$):
\begin{equation}
G^{AB}_{\phi}(p)=\delta^{AB}G_{\phi}(p)={\delta^{AB}}{p^{-2}}
\ \ \mbox{and} \ \ \ G_{\psi}(p)=
{b}M^{2\Delta}C(\Delta){p^{-2(\mu-2+\Delta)}}
\label{propP}
\end{equation}
in the momentum space and
\begin{equation}
G_{\phi}(x)=A{x^{-2\al}}
\ \ \mbox{and} \ \ \ G_{\psi}(x)=
{B}M^{2\Delta}C(\Delta){x^{-2(\bet-\Delta)}}
\label{propX}
\end{equation}
in the coordinate space. Here
$ \al=\mu-1 $,
$ \bet=2 $ and the explicit expressions for the amplitudes
$ A,B $ can be easily deduced from the Eq.~(\ref{propP}).

The renormalization of this theory carried out in the
standard way. The divergences appears as $ \Delta $ - pole
terms in diagrams and can be excluded
by the standard ${\cal R}$ - operation.
The power counting shows that the theory with action~(\ref{action}) is
renormalizable in the \Ne\  in the whole range
$ 2<D<4$ and the renormalized action has the form:
\begin{equation}
S_{R} = -\frac{Z_{1}}{2}(\df\phi)^{2}
 -\frac{1}{2}M^{-2\Delta}\psi K_{\Delta} \psi
 +\frac{Z_{2}}{2}\psi\phi^{2}
 +\frac{1}{2}\psi K \psi.
\label{actionR}
\end{equation}
As distinct from the more usual field theoretic models, the theory under
consideration is not multiplicatively renormalizable. Indeed, the
renormalized action~(\ref{actionR}) cannot been brought into the form of
that~(\ref{action}) by a redefinition of the fields (There is not a
parameter like the coupling constant in~(\ref{action})).
Due to nonmultiplicativity, one fails to apply
\RG\ methods for the
calculation of  critical exponents and has to seek for another
approaches. One of them, which is briefly sketched below relies on the
property of the scale invariance of the theory.  The latter can be
deduced from the equivalence of the nonlinear sigma model to the
$ ({\vec\Phi}^{2})^{2} $ theory in
$ D=4-\eps $ dimensions at the critical point. The more formal proof not
appealing to such equivalence can be found in Ref.~\cite{Nal}.
Now we shall discuss the basic points of the approach mentioned above.
(For more details see Refs.\cite{Step,GraceyA,SS}.)

Let us consider the 1-irreducible Green function $\Gamma(p)$
that depends on the one momentum $p$ only~---~the inverse propagator of
$ \phi $ field or the vertex function with one zero momentum.
We prefer to work with the dimensionless object:
$\bar \Gamma(p) = p^{-d} \Gamma(p)$ ,
$d$ being the canonical dimension of the function $\Gamma(p)$.
Due to
the scale invariance, the form of the
renormalized Green function is fully determined (up to some constant
factor $\bar \Gamma$) by its anomalous dimension
$ \gamma $:
\begin{equation}
\bar \Gamma(p) = \bar \Gamma \cdot
\biggl(\frac{p}{M}\biggr)^{\gamma}.
\label{fix}
\end{equation}
The differentiation of the Eq.(\ref{fix}) with respect
$ M $ yields the equation on the anomalous dimension
\begin{equation}
M \partial_{M} \bar \Gamma(p) = - \gamma \bar \Gamma(p).
\label{scale}
\end{equation}
To calculate the left side of the Eq.(\ref{scale}) we notice
that the all dependence from
$ M $ in the diagrams contributing to
$ \bar \Gamma(p) $ results from the propagators of
$ \psi $  field~(see Eq.(\ref{propP})), i.e. the every diagram acquires
the factor
$M^{2\Delta n}$, where $n$ is the number of $\psi$-lines in a diagram.
Hence, the formula~(\ref{scale})
takes the form:
\begin{equation}
\label{M}
\gamma \bar \Gamma(p) = -2 \Delta
\sum_{\{\Gamma_i\}} n_i \Gamma_i,
\end{equation}
where $n_i$ is the number
of $\psi$-lines in diagram $\bar \Gamma_i$.  The sum runs over all
diagrams $\Gamma_i$ for the renormalized Green function $\bar \Gamma(p)$
and, of course, the limit $\Delta \rightarrow 0$ is implied
in Eq.~(\ref{M}).

\begin{figure}
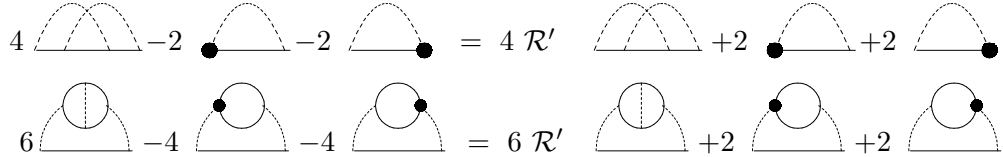

\Lengthunit=0.4cm

\GRAPH(hsize=10){
\mov(0.5,0){\mov(-0.5,0){$4$}\mov(0.4,0){\lin(3.5,0)}
\dasharcto(2.5,0)[+2.5]\mov(1,0){\dasharcto(2.5,0)[+2.5]}}
\mov(6,0){\mov(-1.8,0){$-2$}\mov(0.5,0){\lin(2.5,0)}
\dasharcto(2.5,0)[+2.5]\Circle**(0.5)}
\mov(10,0){
\mov(-1.5,0){$-2$}\mov(0.4,0){\lin(2.5,0)}
\dasharcto(2.5,0)[+2.5]\mov(2.5,0){\Circle**(0.5)}}
\mov(14,0){$=$}
\mov(17,0){\mov(0.5,0){\mov(-2.5,0){$4 \  {\cal R}^{\prime}$}
\mov(0.4,0){\lin(3.5,0)}
\dasharcto(2.5,0)[+2.5]\mov(1,0){\dasharcto(2.5,0)[+2.5]}}
\mov(6.5,0){\mov(-1.8,0){$+2$}\mov(0.5,0){\lin(2.5,0)}
\dasharcto(2.5,0)[+2.5]\Circle**(0.5)}
\mov(10.5,0){
\mov(-1.5,0){$+2$}\mov(0.4,0){\lin(2.5,0)}
\dasharcto(2.5,0)[+2.5]\mov(2.5,0){\Circle**(0.5)}}}}

\vspace{0.3cm}
\Lengthunit=0.3cm

\GRAPH(hsize=10){
\mov(0.5,0){
\mov(-0.5,0){$6$}\mov(0.5,0){\lin(4,0)}
\dasharcto(1,2)[+0.5]\mov(2,2){\mov(0,1){\dashlin(0,-2)}\Circle(2)}
\mov(2.5,2){\dasharcto(1,-2)[+0.5]}}
\mov(7,0){
\mov(-2,0){$-4$}\mov(0.5,0){\lin(4,0)}
\dasharcto(1,2)[+0.5]\mov(2,2){\mov(-1,0){\Circle**(0.5)}\Circle(2)}
\mov(2.5,2){\dasharcto(1,-2)[+0.5]}}
\mov(13.5,0){
\mov(-2,0){$-4$}\mov(0.5,0){\lin(4,0)}
\dasharcto(1,2)[+0.5]\mov(2,2){\mov(1,0){\Circle**(0.5)}\Circle(2)}
\mov(2.5,2){\dasharcto(1,-2)[+0.5]}}
\mov(19,0){$=$}
\mov(22.5,0){
\mov(0.5,0){
\mov(-3.5,0){$6 \  {\cal R}^{\prime}$}\mov(0.5,0){\lin(4,0)}
\dasharcto(1,2)[+0.5]\mov(2,2){\mov(0,1){\dashlin(0,-2)}\Circle(2)}
\mov(2.5,2){\dasharcto(1,-2)[+0.5]}}
\mov(7,0){
\mov(-2,0){$+2$}\mov(0.5,0){\lin(4,0)}
\dasharcto(1,2)[+0.5]\mov(2,2){\mov(-1,0){\Circle**(0.5)}\Circle(2)}
\mov(2.5,2){\dasharcto(1,-2)[+0.5]}}
\mov(13.5,0){
\mov(-2,0){$+2$}\mov(0.5,0){\lin(4,0)}
\dasharcto(1,2)[+0.5]\mov(2,2){\mov(1,0){\Circle**(0.5)}\Circle(2)}
\mov(2.5,2){\dasharcto(1,-2)[+0.5]}}}}
\vspace{0.5cm}
\caption{The example of the rearrangement of the diagrams.
The black dots denote the vertex counterterm. The symbol
${\cal R}^{\prime} $ stands for the standard operation of the
subtracting
of the subdivergencies from a diagram.}
\label{Fr}
\end{figure}

In the first order of
$ 1/N $ expansion Eq.~(\ref{M}) becomes very simple. Indeed,
taking into account that the first order diagrams can develop the simple
$ \Delta $ poles only,
$ \Gamma_i = {(\Gamma_i)_1 }/{\Delta} + (\Gamma_i)_0 + ... $, and
that
$ \bar \Gamma=1 +O(1/N) $ one gets
\begin{equation}
\label{M1}
\gamma = -2 \sum_{\{\Gamma_i\}} n_i (\Gamma_i)_1+O(1/N^{2}).
\end{equation}
Thus for the calculation of the anomalous dimensions in the lowest order
one need know the pole parts of the corresponding diagrams only.
At first sight this property is lost in the next order.  Indeed, the
evaluating of the
$ \gamma $ with \NN\ accuracy requires the knowledge of the \N order
corrections to the constant
$ \bar\Gamma $, which are determined by the finite parts of the \N order
diagrams. This circumstance greatly reduces the effectiveness
of the Eq.~(\ref{M}) for the high order calculations. To avoid this
difficulty one may try to pick out of the sum in the Eq.~(\ref{M})  the
diagrams reproducing the undesirable terms on the lhs of the
latter. Taking in mind the analogous formulae in the dimensional
regularization scheme we regroup the terms in the sum~(\ref{M}) in the
way shown on the Fig~\ref{Fr}. It can be checked,
in these particular cases, that only the
$ {\cal R}^{\prime} $~--~terms contribute to the anomalous dimensions,
whereas the others ensure the cancellation of the extra terms on the lhs
of the Eq.~(\ref{M}). (The $ {\cal R}^{\prime} $  is the standard
operation of the  subtracting of the subdivergencies from diagram.)
Eventually, the formula for the anomalous dimensions
reads
\begin{equation}
\label{MKR}
\gamma = -2 \sum_{\{\Gamma_i\}} n_i ({\cal R}^{\prime} \Gamma_i)_1
+O(1/N^{3}).
\end{equation}
Here the sum runs over all   diagrams (up to \NN order) generated from
the
action~(\ref{action}), i.e. the
nonrenormalized diagrams.
The notation $(f)_{1}$ is used for the residue at the
simple pole in  Laurent expansion of  a function $ f $.
The formula~(\ref{MKR})  gives the simple algorithm
for the calculating of the critical indeces
$ \eta $  and
$ \chi $ defined as
\begin{equation}
\gamma_{\phi}=\eta/2 \ \ \ \mbox{and}\ \ \ \ \gamma_{\psi}=-\eta-\chi
\end{equation}
 up to
 \NN order:
\begin{equation}
\eta=2\sum n_{i}({\cal R^{\prime}} \Gamma_{\phi\phi,i})_{1}, \ \ \ \ \
\chi=-2\sum n_{i}({\cal R^{\prime}} \Gamma_{\phi\phi\psi,i})_{1}.
\label{chieta}
\end{equation}
The advantages of the formula~(\ref{MKR}) for the practical calculations
in the comparison with those~(\ref{M}) is
self-evident. Our next purpose is to derive the analogous formulae in
the
case of arbitrary composite operator. However, to do this we shall take
the advantage of the another approach.

\setcounter{equation}{0}
\section{The extended model}
\label{extend}

Let us consider the model based on the action   obtained
from~(\ref{action}) by introducing two new independent couplings
$ u $  and
$ v $:
\begin{equation}
S = -\frac{1}{2}(\df\phi^{A})^{2}
 -\frac{u}{2}M^{-2\Delta}\psi K_{\Delta} \psi
 +\frac{1}{2}\psi\phi^{2}
 +\frac{v}{2}\psi K \psi.
\label{UVaction}
\end{equation}
In the following we will refer to it as
$ UV $ model. It had been firstly introduced and investigated in
Ref.~\cite{Nal}. Obviously, the initial model~(\ref{action}) is
recovered by the special choice of parameters
$ u=v=1 $.
As distinct from the ordinary sigma model the propagator of
$ \psi $ field in the
$ UV $ model has the more complicated form. Indeed, at
$ v\neq1 $ the last term in the action~(\ref{UVaction})
does not ensure the  exact cancellation of the simple
$ \phi $ loop insertion in
$ \psi  $ line and one should sum up all  such insertions~(see
Fig.~\ref{ins}),
that yields for the propagator of
$ \psi $ field:
\begin{equation}
\label{line}
G_{\psi}(p;u,v) \equiv \frac{1}{u} G_{\psi}(p)\left (1+\frac{v-1}{u}
t^{\Delta}+\frac{(v-1)^2}{u^{2}}
t^{2\Delta} +\ldots\right).
\end{equation}
Here
$ t\equiv C(\Delta)(M^{2}/p^{2}) $.
\begin{figure}
\Lengthunit=0.7cm

\GRAPH(hsize=10){\mov(1.5,0){\mov(-0.6,-0.15)
{$\frac{\sst 1}{\sst u}$}\dashlin(1.5,0) \mov(1.6,-0.15){$+\
\frac{\sst
1-v}{\sst u^2}$} \mov(16.8,-0.15){$+\ ...$}
\mov(5.1,0){\mov(0.95,0){\dashlin(1,0)}\mov(-0.6,0){\dashlin(-1,0)}
\Circle(1.5)}
\mov(7.2,-0.15){$+\  \frac{\sst (1-v)^2}{\sst u^3}$}
\mov(11.4,0){\mov(0.95,0){\dashlin(1,0)}\mov(-0.6,0){\dashlin(-1,0)}
\Circle(1.5)}
\mov(13.9,0){\mov(0.8,0){\dashlin(1,0)}\Circle(1.5)}}}
\caption{ The effective $\psi$ - line }
\label{ins}
\end{figure}
The theory is evidently
renormalizable and the renormalized action takes form:
\begin{equation}
S_{R} = -\frac{Z_{1}}{2}(\df\phi^{A})^{2}
 -\frac{u}{2}M^{-2\Delta}\psi K_{\Delta} \psi
 +\frac{Z_{2}}{2}\psi\phi^{2}
 +\frac{v}{2}\psi K \psi,
\label{UVRaction}
\end{equation}
where  the renormalization constants
 depend on the coupling
$ u,v $ also ( $ Z_{i}=Z_{i}(u,v,.. ) $).
The redefinition of the fields and the couplings
\begin{equation}
\label{bare}
\Phi_{0}=Z_{\Phi}\Phi,\ \
u_0=u M^{-2\Delta} Z_u,\ \ v_0=v Z_v,
\end{equation}
where
$ \Phi\equiv \{\phi,\psi\} $,
$ Z_{\phi}=Z_{1}^{1/2} $,
$ Z_{\psi}=Z_{2}Z_{1}^{-1} $ and
$ Z_{u}=Z_{v}=Z_{\psi}^{-2} $,  brings the action~(\ref{UVRaction}) into
the initial form, hence, the
$ UV $ model is the multiplicatively renormalizable.  This implies that
the model can be analyzed by the standard \RG\ methods. The basic \RG\
functions are defined as follows:
\begin{equation}
\gamma_{\Phi}=M\df_{M}\ln{Z_{\Phi}}, \ \
\bet_{u}=M\df_{M}u=2\Delta u-2u\gamma_{\psi},\ \
\bet_{v}=M\df_{M}v=-2v\gamma_{\psi}.
\end{equation}
The \RG\ equation for the one particle irreducible ($ 1PI $)
Green functions reads as:
\begin{equation}
(M\df_{M}+\bet_{U}\df_U-n_{\Phi}\gamma_{\Phi})
\Gamma(p_{1},\ldots,p_{n})=0.
\label{RG1}
\end{equation}
Here we used the shorthand  notations
$ U=\{u,v\} $,
$ \bet_{U}\df_U =\bet_{u}\df_u+\bet_{v}\df_v $ and
$n_{\Phi}\gamma_{\Phi}=n_{\phi}\gamma_{\phi} +n_{\psi}\gamma_{\psi} $.
For the set of the composite operators
$ \{ F_{i}\} $ mixing under renormalization the corresponding equation
has the matrix form:
\begin{equation}
\left(\left [
M\df_{M}+\bet_{U}\df_U-n_{\Phi}\gamma_{\Phi}\right ]\delta^{ik}
+\gamma_{F}^{ik}\right)
\Gamma_{k}(p;p_{1},\ldots,p_{n})=0.
\label{RG2}
\end{equation}
Here $ \Gamma_{i}(p;p_{1},\ldots,p_{n})$ is the
$ n $-points
$ 1PI $ Green function with insertion of the renormalized operator
$[F_{i}]_{R}  $. The latter is defined as
$
[F_{i}]_{R}= Q_{ik}F_{k}
$
and
$ Q_{ik} $ is chosen to ensure finiteness of all Green functions of
operator $ [F_{i}]_{R}$.
As usual, the summation over repeated indexes is implied.

At last the matrix of anomalous dimensions
$ \gamma_{F} $ is defined as
\begin{equation}
  \gamma_{F}=-M\df_{M}Z_{F}Z_{F}^{-1},\ \ \
Z^{ik}_{F}=Q^{ik}Z_{\Phi}^{-n_{\Phi,k}}.
\end{equation}
Here $Z_{\Phi}^{-n_{\Phi,k}}\equiv
Z_{\phi}^{-n_{\phi,k}}Z_{\psi}^{-n_{\psi,k}}$ and the multiindex
$n_{\Phi,k}\equiv \{n_{\phi,k},n_{\psi,k}\} $ shows the number of the
$ \phi $ and
$ \psi $ fields in the monomial
$ F_{k} $.

We remind, that our purpose is the calculation of the critical
indices of composite operators in the ordinary sigma model, i.e. we are
interested in the limit
$ u,v\to 1 $ of the
$ UV $ model. However, in this limit the beta functions
$ \bet_{u},\bet_{v} $ are not zero. (Indeed,
$ \bet_{u}=\bet_{v}=\gamma_{\psi}\neq0 $.) Thus, generally speaking the
point $ u=v=1 $ is not the critical point of $ UV $ model and the \RG\
functions $ \gamma_{\Phi}, \gamma_{F} $ calculated at
$ u=v=1 $ are not coincide with the anomalous dimensions of the fields
(operators) in the ordinary sigma model.
However, it had been shown in Ref.~\cite{Nal} that at the particular
choice of the renormalization scheme (the so called scheme of
subtraction
at fixed momenta, we refer to it as the scheme I) the renormalized Green
functions depend on the difference $  u -  v $ only.
In this case the anomalous term
\begin{equation}
\label{A}
{\cal A} \equiv -\bet_{ U}\df_{ U}
{\Gamma}^I_{i}(u-v;p,p_{1},\ldots,p_{n})\bigl |_{u=v=1}
\end{equation}
is dropped out of the Eq.~(\ref{RG2})
and the \RG\ functions  $ \gamma_{\Phi}, \gamma_{F} $ calculated in this
scheme are the true anomalous dimensions.
The Green functions calculated in  any other renormalization
scheme are related to those
${ \Gamma}_{i}^{I}(\bar u- \bar v,...)$
calculated in the scheme~I by a finite
renormalization:
\begin{equation}
{ \Gamma}_{i}(u,v,...)=Z_{ik}^{*}(u,v)\cdot
{ \Gamma}_{k}^{I}(\bar u-\bar v,...).
\label{renorm}
\end{equation}
Here
$\bar u = Z^{*}_u(u,v)\cdot u \ ,\ \bar v = Z^{*}_v(u,v)\cdot v$
and
$Z_{ik}^{*}(u,v),\ Z^{*}_u(u,v),\ Z^{*}_v(u,v)$
are the constants of the finite renormalization.
Taking into account that the bare couplings
$ u_{0}, v_{0}$
are independent from the
renormalization scheme, we obtain from the Eq.~(\ref{bare})
that $u/v = \bar u/ \bar v$ and, hence,
$ Z^{*}_u=Z^{*}_v$.
Then, from the Eq.~(\ref{renorm}), one
immediately obtains      that Green functions
calculated in the different regularization schemes at $u=v=1$
differ one from
another only by a constant multiplier.
This proves
that the critical exponents calculated in
the model based on action~(\ref{action}) are independent from
 the regularization scheme being used and, in particularly, from
the choice of the function $C(\Delta)$.

Further, inserting
$ \Gamma_{i} $ in the form~(\ref{renorm}) into Eq.~(\ref{RG2}) one
obtains the
following equation for the Green functions of the ordinary sigma model:
\begin{equation}
\left(\left [
M\df_{M}-n_{\Phi}\gamma_{\Phi}\right ]\delta^{ik}
+(\gamma_{F}^{ik}+\gamma_{F}^{*,ik})\right)
\Gamma_{k}(p;p_{1},\ldots,p_{n})=0.
\end{equation}
where
$\gamma_{F}^{*}=
(\bet_{u}\df_u+\bet_{v}\df_v) Z^{*}(u,v)Z^{*-1}(u,v)|_{u=v=1}  $.
Thus the anomalous dimensions are given by the eigenvalues of the matrix
$\gamma_{F}+\gamma_{F}^{*}   $, but not of the
$  \gamma_{F}$ alone.
Of course, the
$ \gamma_{F}^{*}  $  is zero in the scheme I, but the latter is not very
suitable for the practical calculations. The most convenient scheme for
the calculations of \RG\ functions is the \MS\ scheme, but then one has
to know  $ \gamma_{F}^{*}  $.
Fortunately, the matrix $ \gamma_{F}^{*}  $
is appeared to be of
$ O(1/N^{3}) $ in the \MS\ scheme, i.e. for the determination of the
 critical exponents up to $ 1/N^{2} $ order it is sufficient to know $
\gamma_{F}$ only.  Henceforth we restrict ourselves to \MS\ scheme only.
The \RG\ functions
$ \gamma $, then, are related to the simple
$ \Delta $ poles of the corresponding renormalization constants:
\begin{equation}
\label{RG}
\gamma_{\psi}=\left. 2 \df_{u} Z^{(1)}_{\psi} \right |_{u=v=1},
\ \
\gamma_{\phi}= \left. 2 \df_{u} Z^{(1)}_{\phi} \right |_{u=v=1}, \ \
\gamma^{ik}_{F}=\left. -2 \df_{u}  Q_{ik}^{(1)} \right |_{u=v=1} +
\delta_{ik}n_{\Phi,k} \gamma_{\Phi},
\end{equation}
where
$ Z_{i}(u,v)=1+\sum_{n=1}^{\infty}Z_{i}^{(n)}(u,v)/\Delta^{n} $,\ \
$ Q_{ik}(u,v)=\delta_{ik}+
\sum_{n=1}^{\infty}Q_{ik}^{(n)}(u,v)/\Delta^{n} $.
The mixing matrix $ Q_{ik} $ is calculated directly from the diagrams
as the divergent part of the  functional
$\Gamma_{F_i}(x;\Phi)$:
\begin{eqnarray}
&&{{\cal KR}'} \Gamma_{F_i}(x;\Phi) =-
\sum_{n=1}^{\infty} \frac{1}{\Delta^n} \cdot Q_{ik}^{(n)} F_k (x),
\label{GF}  \\
&&\Gamma_{F_i}(x;\Phi) \equiv  \sum_{n} \frac{1}{n!}
\int {\rm d}x_1...{\rm d}x_n \Gamma_{F_i,n}(x;x_1...x_n)
\Phi(x_1)...\Phi(x_n). \nonumber
\end{eqnarray}
Here $ \Gamma_{F_i}(x;x_{1},\ldots,x_{n})$ is the nonrenormalized
(i.e. one arising at averaging with action~(\ref{UVaction}))
$ n $-points $ 1PI $ Green function with operator $F_{i}$
insertion
and the operation ${\cal K}$  selects the
 singular ($\Delta$ poles)  part of the diagrams.

Because there are not derivatives with respect
$ v $ in the formulae~(\ref{RG}), we can put
$ v=1 $ from the very beginning. In this case the propagator of
$ \psi $ field~(\ref{line}) is reduced up to factor
$ 1/u $ to that in the ordinary sigma model:
$ G_{\psi}(p,u,v=1)=G_{\psi}(p)/u $.
Then taking into account that the constants
$ Z_{i}, Q_{ik} $  are determined by the divergent (pole) parts of the
corresponding Green functions and  that operation
$\df_u $ gives simply the factor $ -n_{\psi} $ ($ n_{\psi} $ is the
number of
$ \psi $ lines) for each diagram, one concludes that all calculations
needed for determining of anomalous dimensions~(\ref{RG}) can be done
in the framework of the standard $ \sigma $ model.

Let us prove now that matrix
$ \gamma^{
*}_{F} $   vanishes in the
$ 1/N^{2} $ order. For this it is sufficient to show that the anomalous
term
$ \cal A $~(see Eq.~(\ref{A})) is zero in this order. First of all,
note, $ \gamma_{\psi} $ being of
$ O(1/N) $ order, we need to prove that the remaining part is of
$ O(1/N^{2}) $ order.
Let us consider the arbitrary first order diagram
of the ordinary
$ \sigma $ model   with
$ n $  internal
$ \psi $ lines.  Regarding the regulator
$ \Delta $ on each line as the independent variable and picking out the
multiplier
$ C(\Delta) $ from the
$ \psi $  propagator we write the answer for the diagram as
$ C^{n}(\Delta)G(\Delta,\ldots,\Delta) $. Then, taking into account the
form of the
$ \psi $ propagator~Eq.~(\ref{line}) one obtains the following
expression for this diagram in the extended model:
\begin{equation}
 \frac{1}{u^{n}}\sum_{m_{1},\ldots,m_{n}=1}^{\infty}
(C(\Delta))^{m_{1}+\ldots+m_{n}}
G(m_{1}\Delta,\ldots,m_{n}\Delta)
(\frac{v-1}{u})^{m_{1}+\ldots+m_{n}-n}.
\label{full}
 \end{equation}
As usual, the renormalized diagram is obtained from the nonrenormalized
one by the
$ {\cal R} $ operation. In the Ref.~\cite{Nal} it was shown that the
${\cal R} $ operation based on the subtraction scheme I gives for each
term in the sum~(\ref{full}) the same (up to factor
$((v-1)/u)^{...}   $)
answer. (Then, the sum~(\ref{full}) becomes a trivial and results in the
factor $ (1+u-v)^{-n} $.) The above property  holds no longer in the
\MS\ scheme. The action of the
$ {\cal R} $ operation on different terms in the sum~(\ref{full}) leads
to the essentially different answer even in the first order of \N\
expansion.  However, in the first order of the \N\ expansion the whole
sum~(\ref{full}) after renormalization depends on the difference $ u-v $
only too.

In the case of a convergent diagram
$ G $ this statement is trivial. Let us consider the superficially
divergent diagram. In the \N\ order such diagram having no
subdivergencies, we represent function
$ G(\ldots) $ as:
\begin{equation}
G(m_{1}\Delta,\ldots,m_{n}\Delta)=
\frac{F(m_{1}\Delta,\ldots,m_{n}\Delta)}{(m_{1}+\ldots+m_{n})\Delta},
\label{repr}
\end{equation}
where the function
$ F $  is regular in the neighborhood of zero in all its arguments.
Since in the \MS\ scheme the
$ R $ operation is reduced to the removing of the
$ \Delta $ poles one has for the renormalized diagram:
\begin{equation}
 \frac{1}{u^{n}}\sum_{m_{1},\ldots,m_{n}=1}^{\infty}
\left( C^{\prime}(0)F(0)+
\frac{m_{1}F_{1}^{\prime}(0)+
\ldots+m_{n}F_{n}^{\prime}(0)}{m_{1}+\ldots+m_{n}}
\right)
(\frac{v-1}{u})^{m_{1}+\ldots+m_{n}-n}.
\label{Sum1}
\end{equation}
 Taking into account that the terms like
$ m_{1}/(m_{1}+\ldots+m_{n}) $  due to the evident symmetry of
the rest of summand  in
$ m_{i} $  can be replaced by
$ 1/n $,
one obtains for the sum~(\ref{Sum1})
\begin{equation}
 \frac{nC^{\prime}(0)F(0)+ F_{1}^{\prime}(0)+
\ldots+F_{n}^{\prime}(0)}{n(1+u-v)^{n}}
\end{equation}
The case when a superficially convergent diagram has a divergent
subgraph can be treated along these lines as well.

Since the anomalous term
$ {\cal A} $ is proportional  to
$ (\df_u+\df_{v}) \Gamma(p_{i},u,v)|_{u=v=1}$   we conclude that
$ \cal A $, and hence,
$ \gamma^{*} $ vanish up to \NN\ order.
(Note also, that we proved the
more strong statement $ A(u,v)=0 $ at
$ u=v $, than it was needed: $ A(u,v)=0 $ at $ u=v=1 $.)
Unfortunately, the arguments used above do not work
in the next orders of \N expansion, since
the diagrams    which
give the nonzero contributions to the anomalous term
$ \cal A $ can be easy found in the higher orders.
Though we cannot to reject the possibility of the cancellation of
the contributions from the different diagrams, it seems to be very
unlikely.

Thus we have shown
that up to \NN order the anomalous dimensions of composite
operators are simply related to the
$ \Delta $ poles of the corresponding diagrams. The simplicity of the
formulae~(\ref{RG}) in the combination with the uniqueness of the triple
vertex (at $ \Delta=0 $) and masslessness of propagators makes the \NN\
order calculations a rather straightforward task. The only thing which
causes some problem is a large number of diagrams to be evaluated. In
the next section we show how to overcome this difficulty.

\setcounter{equation}{0}

\section{Conformal structures in perturbation theory.}
\label{confors}

\begin{figure}
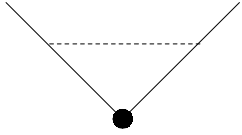

\Lengthunit=0.5cm
\centerline{
\GRAPH(hsize=10){
\mov(3,0){\Circle**(0.5)\lin(-3.1,3.1)\lin(3.1,3.1)
\mov(-2.2,2){\dashlin(4,0)}}}}
\caption{ The $1/N$ -  diagrams for the
Green function with the insertion of the operator $\phi_{A}\phi_{B}$. }
\label{FN1}
\end{figure}

\begin{figure}
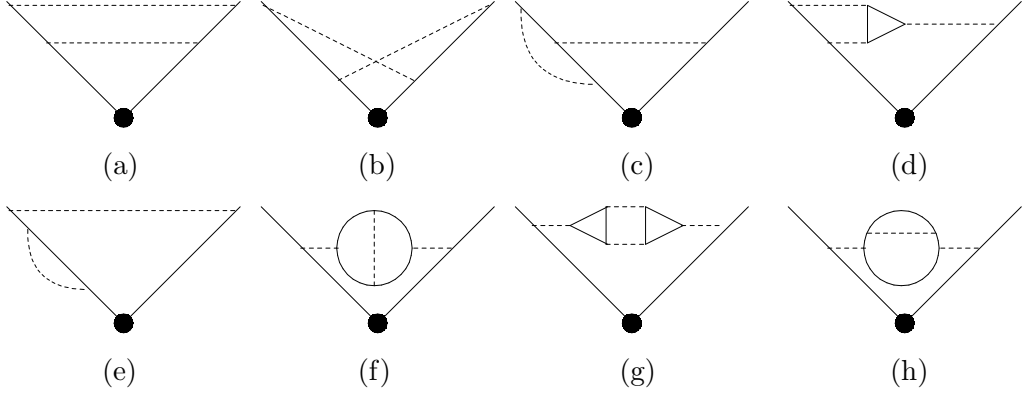

\Lengthunit=0.5cm
\GRAPH(hsize=10){
\mov(3,0){\Circle**(0.5)\lin(-3.1,3.1)\lin(3.1,3.1)
\mov(-3.3,3){\dashlin(6,0)}\mov(-2.3,2){\dashlin(4,0)}}
\mov(9.5,0){\Circle**(0.5)\lin(-3.1,3.1)\lin(3.1,3.1)
\mov(-3.3,3){\dashlin(4,-2)}\mov(2.7,3){\dashlin(-4,-2)}}
\mov(16,0){\Circle**(0.5)\lin(-3.1,3.1)\lin(3.1,3.1)
\mov(-2.3,2){\dashlin(4,0)}\mov(-1.3,0.9){\dasharcto(-1.9,2)[+1]}}
\mov(23,0){\Circle**(0.5)\lin(-3.1,3.1)\lin(3.1,3.1)
\mov(-3.3,3){\dashlin(2,0)}\mov(-2.3,2){\dashlin(1,0)}
\mov(-1.5,3){\lin(0,-1)\lin(1,-0.5)\mov(0,-1){\lin(1,0.5)}
\mov(0.8,-0.5){\dashlin(2.3,0)}}}
\mov(1.4,-1,5){(a)}\mov(8.2,-1,5){(b)}
\mov(14.9,-1,5){(c)}\mov(22.1,-1,5){(d)}}
\vspace{0.3cm}
\GRAPH(hsize=10){
\mov(3,0){\Circle**(0.5)\lin(-3.1,3.1)\lin(3.1,3.1)
\mov(-3.3,3){\dashlin(6,0)}\mov(-1.3,0.9){\dasharcto(-1.5,1.6)[+1]}}
\mov(9.5,0){\Circle**(0.5)\lin(-3.1,3.1)\lin(3.1,3.1)
\mov(-2.3,2){\dashlin(1,0)\mov(3,0){\dashlin(1,0)}
\mov(-0.3,0){\mov(2,0){\Circle(2)}\mov(2,1){\dashlin(0,-2)}}}}
\mov(16,0){\Circle**(0.5)\lin(-3.1,3.1)\lin(3.1,3.1)
\mov(-2.9,2.6){\dashlin(1,0)\mov(1,0){\lin(1,0.5)\lin(1,-0.5)
\mov(0.7,0.5){\lin(0,-1)\dashlin(1,0)\mov(0,-1){\dashlin(1,0)}}
\mov(2.5,0){\lin(-1,0.5)\lin(-1,-0.5)\mov(-1,0.5){\lin(0,-1)}}
\mov(2.2,0){\dashlin(1,0)}}}}
\mov(23,0){\Circle**(0.5)\lin(-3.1,3.1)\lin(3.1,3.1)
\mov(-2.3,2){\dashlin(1,0)\mov(3,0){\dashlin(1,0)}
\mov(-0.3,0){\mov(2,0){\Circle(2)}\mov(1.1,0.4){\dashlin(1.8,0)}}}}
\mov(1.4,-1,5){(e)}\mov(8.2,-1,5){(f)}
\mov(14.9,-1,5){(g)}\mov(22.1,-1,5){(h)}}
\caption{ The $1/N^2$ -  diagrams for the
Green function with insertion of operator $\phi_{A}\phi_{B}$. }
\label{FN2}
\end{figure}

Everyone being a familiar with the \N expansion knows that the number of
diagrams to be evaluated increases drastically with a order of
expansion. To receive the impression about the rate  of the growth it is
sufficient to  compare the number of the first and the second order
diagram for any operator. Indeed, one has only one \N order
diagram~(Fig.~\ref{FN1})
 and
eight ones in \NN order~(Fig.~\ref{FN2}) for the \AB\ operator. The
analogous comparison for the $ \psi^{2} $ operator gives  3 and 69
diagrams, respectively.  However, it can be easy realized that the bulk
of the second order diagrams are the nothing else as the first order
ones with the insertion of the vertex or the propagator corrections.
It is intuitively clear that taking into account of these corrections
must lead to the dressing of the bare propagators and vertices.
In what follows we
shall show that this really occurs.  For further discussion  to be  more
transparent we will use \AB\ operator as illustrative example.  The
second order diagrams for the two-point Green function
with operator \AB\
insertion are drawn on Fig.~\ref{FN2}.  Two of them~(c,d)
arise due to the vertex corrections to the first order
diagram~(Fig.~\ref{FN1}); one~(e)~--~due to the self-energy
(SE) insertion in $ \phi $ line and three~(f,g,h)~--~due to
the SE insertions in $ \psi $ line.

We are interested in the contributions from the diagrams of this
type~(c~--~g on Fig.~\ref{FN2}) to the matrix
$ \df_u Q^{ik}|_{u=1} $. The more precisely we need the simple
$ \Delta $ pole term in the sum
\begin{equation}
\sum_{\{\Gamma_{i}\}} n_{i}\left ({\cal R'} \Gamma_{i}\right).
\label{sum1}
\end{equation}

\vspace*{0.3cm}

\noindent
Here sum runs over the \NN\ order nonrenormalized diagrams
$\{\Gamma_{i}\}   $ which arise
from the \N\ order~--~base~--~diagrams after taken into account the
vertex and the propagator corrections~(the diagrams~c-h on the
Fig.~\ref{FN2}).
For later convenience we rewrite the sum~(\ref{sum1})
in the following form:
\begin{equation}
\sum_{\{\Gamma_{i}\}} n_{i}\left ({\cal R'} \Gamma_{i}\right)=
\left (\sum_{\{\Gamma_{i}\}} n_{i} \Gamma_{i}-
\sum_{\{\bar\Gamma_{i}\}}\bar n_{i} \bar\Gamma_{i}\right )-
\sum_{\{\bar\Gamma_{i}\}} \wh n_{i} \bar\Gamma_{i}.
\label{sum2}
\end{equation}
Here
$ {\{\bar\Gamma_{i}\}}  $ is the set of the
counterterm diagrams corresponding to the ones from the
set
$\{ \Gamma_{i}\} $, i.e
$ {\cal R^{\prime}}\Gamma_{i}=\Gamma_{i}-\bar \Gamma_{i} $;
$ \bar n_{i} $ is the number
of $ \psi $ lines in the  diagram
$\bar \Gamma_{i} $, while
$ \hat n_{i}=n_{i}-\bar n_{i} $ is the number of
$ \psi $ lines in the divergent subgraph.
It is appeared that
the last sum in the Eq.~(\ref{sum2}) has no simple
$ \Delta $ pole, provided that
 the  relation $2n_{\psi}=n_{\phi}=n_{V} $ between the
number of the internal lines and the triple vertices
holds for the \N\ order base diagram.
Indeed, one can see that the divergent subgraphs in
the diagrams with
the $ SE $ insertions in
$ \psi $ line (Fig.~\ref{FN2}~f--h) are the same as those in the
diagrams with the vertex corrections and  the
$ SE $  insertion in
$ \phi $ line (Fig.~\ref{FN2}~c--e).
We combine the diagrams into three different
groups~---~(c,f), (d,g), (e,h)~--~ in accordance
with the type of the
divergent subgraph.
Obviously, the diagrams appearing
after reducing of the divergent subgraph in the diagrams from the same
group, i.e. those from the set
$ \{\bar\Gamma_{i}\} $,  differ one from another only by the
simple $ \phi $ loop insertion in $ \psi  $ line. For example, for $
S/\Delta $ being the pole part of the divergent subgraph, the
contribution of the (e,h) pair into the sum in question
reads:
\vskip 0.3cm

\Lengthunit=0.4cm
\GRAPH(hsize=10){
\mov(3.8,1.5){$\frac{\textstyle S}{\textstyle \Delta}\times\Biggl ($}
\mov(10,0){\Circle**(0.5)\lin(-3.1,3.1)\lin(3.1,3.1)
\mov(-2.3,2){\dashlin(4,0)}}
\mov(14.0,1.5){$+\ \ \ \frac{\textstyle 1}{\textstyle 2}$}
\mov(19,0){\Circle**(0.5)\lin(-3.1,3.1)\lin(3.1,3.1)
\mov(-2.3,2){\dashlin(1,0)\mov(3,0){\dashlin(1,0)}
\mov(-0.3,0){\mov(2,0){\Circle(2)}}}}
\mov(23,1.5){$\Biggr )$}}
\vskip 0.3cm
(The factor
$ 1/2 $ here is due to the symmetry coefficient of diagram (h).)
The analytical expression for the above sum can be represented as
\begin{equation}
\frac{S}{\Delta}C(\Delta)\left (\frac{F(\Delta)}{\Delta}-
C(\Delta)\frac{F(2\Delta)}{2\Delta}\right ),
\label{example}
\end{equation}
$ F $ being the regular function of
$ \Delta $. (To derive~(\ref{example}) it is sufficient to remember
the definitions~(\ref{Kp}),(\ref{propP}),(\ref{propX})).
It is seen the difference~(\ref{example})  does not contain the simple
$ \Delta $ pole. Obviously, the same arguments work and in a general
case, the condition
$ 2n_{\psi}=n_\phi=n_{V} $ ensuring the cancellation of the simple pole
contributions among each group separately. Henceforth we shall only
deal with the diagrams satisfying this condition.  Moreover, we confine
our consideration by those diagrams for which the only subdivergencies
are due to the vertex and SE subgraphs. In spite of that these
conditions restrict the class of the admissible operators it appears to
be wide enough to justify the necessity of the present discussion.

We start with the analyzing of the diagrams with the $SE$ insertions,
say, in  $ \phi $ lines~(Fig.~\ref{FN2}~e).
 For later convenience we will regard the line indices as the variables
and distinguish
 the regulators
$\Delta$ on the $\psi$ - lines being
internal
($ \Delta_{1} $) and external ($ \Delta_{2} $) with respect to the
divergent subgraph.
Also, we introduce
the regular function
$ F $
related to the base diagram~(Fig.~\ref{FN1}) with the line indeces
differing from their canonical values $ \al_{\phi}=1 $  and $
\al_{\psi}=\mu-2 $  by $ \Delta_{i} $ as follows
\begin{equation}
G=\frac{F(\al_{\phi}+\Delta_1,\al_{\phi}+\Delta_2,\al_{\psi}+
\Delta_3)}{\Delta_1+\Delta_{2}+\Delta_3}.
\end{equation}
Then the answer for the diagram with the SE insertion in
$ \phi $ line can be written as
\begin{equation}
\frac{S(\Delta_1)}{\Delta_1}\cdot
\frac{F(\al_{\phi}+\Delta_1,\al_{\phi,}
\al_{\psi}+\Delta_2)}{\Delta_1+\Delta_2},
\end{equation}
where the factor $S(\Delta_{1})/\Delta_{1}$ result from the integration
over loop momenta in the subgraph. For the corresponding counterterm
diagram  one has ${S(0)}/{\Delta_1}\cdot
{F(\al_{\phi},\al_{\phi,}
\al_{\psi}+\Delta_2)}/{\Delta_2}.
$
Their contribution to the sum~(\ref{sum2}) reads
($n=2$  and  ${\bar n}=1$ in this case)
\begin{equation}
2\cdot
\frac{S(\Delta_1)}{\Delta_1}\cdot
\frac{F(\al_{\phi}+\Delta_1,\al_{\phi},
\al_{\psi}+\Delta_2)}{\Delta_1+\Delta_2} -
\frac{S(0)}{\Delta_1}\cdot
\frac{F(\al_{\phi},\al_\phi,\al_{\psi}+\Delta_2)}{\Delta_2}.
\end{equation}
Evaluating of the above difference
at the "symmetric" point $\Delta_1=\Delta_2=\Delta$ yields
for the simple pole residue
\begin{equation}
\label{contr}
S(0)
F_{1}^{\prime}(\al_{\phi},\al_{\phi},\al_{\psi}) +
S^{\prime}(0) F(\al_{\phi},\al_{\phi},\al_{\psi}).
\end{equation}
Noticing that the constant $ -S(0) $  is nothing else as the
anomalous dimension of $ \phi $ field:
$ \gamma_{\phi}=-S(0)+O(1/N^{2}) $ we relate
the above expression
to the base diagram with one  bare
$ \phi $ propagator being replaced by the dressed one:
$ 1/p^{2}\to A/p^{2-\eta} $
($ A=1+S^{\prime}(0) $ and
$ \eta=2\gamma_{\phi} $).  Namely, the expression~(\ref{contr})
can be obtained as the
\NN - order term in the expansion of the following object
\begin{equation}
\label{1}
A \cdot F(\al_{\phi}-\gamma_{\phi},\al_{\phi},\al_{\psi}).
\end{equation}

It is not more hard to see that taking into account the SE insertions
in $\psi$ line also leads to the dressing of this line, i.e. the
contributions of these diagrams to the sum~(\ref{sum2}) can be received
from the expansion of the object:
\begin{equation}
\label{2}
B\cdot F(\al_{\phi},\al_{\phi},\al_{\psi}-\gamma_{\psi}).
\end{equation}

The following necessary step is the inclusion into consideration of the
vertex corrections. Indeed, having restricted by the propagators
corrections only one loses the more attractive feature of \Ne~---~the
uniqueness of the triple vertex.
The analysis of the vertex
corrections is a bit more complicated than those of the
$ SE $ insertions. Indeed, the diagram with insertion in a line has, in
fact, the same topology as the initial one, and hence, (that is
crucial) they both can be described by one and the same function.
To achieve this in the case in question, where
the topology of the diagrams is essentially
different, we proceed in the following way.
\begin{figure}
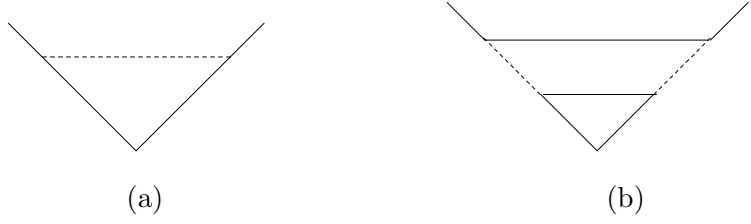

\Lengthunit=0.5cm

\GRAPH(hsize=10){
\mov(8,0){\mov(0,-1.5){(a)}
\lin(-3.4,3.4)\lin(3.4,3.4)
\mov(-2.75,2.5){\dashlin(5,0)}}
\mov(20,0){{\mov(0.5,-1.5){(b)}}
\lin(-1.5,1.5)\lin(1.5,1.5)
\mov(-1.77,1.5){\dashlin(-1.5,1.5)}\mov(1.25,1.5){\dashlin(1.5,1.5)}
\mov(-3.5,2.95){\lin(5.95,0)\lin(-1.0,1.0)\mov(6,0){\lin(1,1)}}
\mov(-2.2,1.5){\lin(3,0)}
}}
\caption{The
$ 1/N $ order vertex diagrams}
\label{VER}
\end{figure}
First of all, let us consider the vertex function itself. Up to terms
of
$ O(1/N^{2}) $ order it reads
\begin{equation}
\gamma_{R}(p,q)= 1+\frac{\gamma_{1}(\Delta;p,q)}{\Delta}
+\frac{\gamma_{2}(\Delta;p,q)}{2\Delta} -\mbox{poles}=
1+\gamma_{1}^{\prime}(p,q)+\frac12\gamma_{2}^{\prime}(p,q).
\label{gR}
\end{equation}
Here
$ \gamma _{1}(\Delta)/\Delta$,
$ \gamma _{2}(\Delta)/2\Delta$  are the contributions of the diagrams
shown on the Fig.~{\ref{VER}}a,b, respectively. (In the further we shall
not show explicitly the momentum dependence in the functions
$ \gamma_{i}, \gamma_{R} $. Note only,  that at
$ \Delta=0 $ the functions
$ \gamma_{i} $ are independent from
$ p,q $.)
Due to conformal invariance of the theory,
the form of the {\it full} vertex
$ \gamma_{R} $ is fixed up to the constant factor. In coordinate space
it reads
\begin{equation}
\gamma_{R}(x,y;z) \equiv
\hat Z (z-x)^{-2\alpha}(z-y)^{-2\alpha}(x-y)^{-2\beta},
\end{equation}
where
$ \al=\mu-1-\gamma_{\psi}/2 $,
$ \bet=2-\gamma_{\phi}+\gamma_{\psi}/2 $.

Further, taking in mind the pole structure of the diagrams (c,d) on
Fig.~{\ref{FN2}} we write for the latter
\begin{equation}
\frac{\wh F[\gamma_{m}(\Delta_1);\Delta_2]}
{m\Delta_1(m\Delta_1+\Delta_2)},
\end{equation}
where
$ m=1,2 $, respectively.
The function
$ {\wh F} $ is  regular   with respect the variables
$ \Delta_{1(2)} $ and is defined as follows :
\begin{equation}
 {\wh F}[\gamma_{m}(\Delta_1);\Delta_{2}]
=(m\Delta_{1}+\Delta_{2})
\int \frac{d^{2\mu}p}{(2\pi)^{2\mu}}
\gamma_{m}(\Delta_{1},p,p+q){ (G/\gamma_{1})}(p,...,\Delta_{2}).
\label{fun}
\end{equation}
Here the function
$(G/\gamma_{1})(p,....\Delta_{2}) $ is related to the diagram
with the vertex subgraph removed.
(For the diagram on Fig.~\ref{FN2}c,d
it is given by the product of three propagators).
The evaluation of the
contribution of diagrams~(c,d) on Fig.~\ref{FN2} together with
the corresponding counterterm diagrams
yields for the simple pole residue in  the sum~(\ref{sum2})
\begin{equation}
 \df_\Delta\left(\wh F[\gamma_1(\Delta);0]+
\frac12\wh F[\gamma_2(\Delta);0]\right)\Bigl |_{\Delta=0}.
\label{vert}
\end{equation}
Looking on the Eq.~(\ref{fun}) one can see that the renormalized  vertex
$ \gamma_{R} $ should appear in the natural way if it were be
possible to  carry out
differentiating with respect
$ \Delta $ under the sign of integral. However, at
$ \Delta_{2}=0 $ the integral evolves the pole in
$ \Delta_{1} $, therefore this operation is not allowed. To obviate
this difficulty, one can, taking into account the regularity of the
function
$ \wh F $ in $ \Delta_{1(2)} $,  carry out
differentiation at the nonzero  second argument $ \Delta_{2} $ and then
take the limit $ \Delta_{2} \to 0$. At first step we obtain
for~(\ref{vert})
\begin{equation}
 \frac{\wh F(\gamma_{1}(0)+\gamma_{2}(0),\Delta_{2})}{\Delta_{2}}+
\Delta_{2}
\int \frac{d^{2\mu}p}{(2\pi)^{2\mu}}
\gamma_{R}^{(1)}(p,p+q){ (G/\gamma_{1})}(p,...,\Delta_{2}),
\label{ver1}
\end{equation}
where
$ \gamma_{R}^{(1)}=\gamma_{1}^{\prime}+
\gamma_{2}^{\prime}/2 $~(see Eq.~(\ref{gR})).
To proceed further we represent $ \gamma_{R}^{(1)}$ as
$ \gamma_{R}^{(1)}=\df_t\gamma_{R}(t/N)|_{t=0} $ and, again, change
the order of the differentiation and integration.
At last, taking into account that scaling dimension
$ \chi$~(\ref{chieta}) of the function $ \gamma_{R} $ is equal to
$-2(\gamma_{1}(0)+\gamma_{2}(0)) $ at first $ 1/N $ order, so that the
integral in~(\ref{ver1}) evolves the pole in $ (\Delta_{2}-t\chi/2) $,
we find eventually for~(\ref{vert}) the following representation
\begin{equation}
\label{3}
\df_{t}\left (-\frac{t\chi}{2} \int
\frac{d^{2\mu}p}{(2\pi)^{2\mu}} \gamma_{R}(t/N,p,p+q)
{(G/\gamma_{1})}(p,...,0) \right)\Bigl |_{t=0}.
\end{equation}
In other
words the contribution of the diagrams in question is obtained from the
\N expansion of the diagram with the dressed vertex.

Basing on these results~(Eqs.~(\ref{1}),(\ref{2}),(\ref{3}))
 we can formulate the simple algorithm for the
calculating of the contribution of the diagrams with the propagator and
vertex corrections to the matrix of the anomalous dimensions.
Let G to be a \N order diagram and the diagrams
$ \{G_{i}\} $  are the \NN order ones from the class under
consideration.
Then to obtain the contribution of the diagrams
$ \{G_{i}\} $ to the anomalous dimension one need:

\begin{figure}
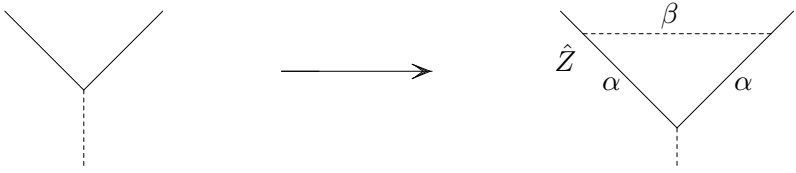

\Lengthunit=0.5cm
\GRAPH(hsize=5)
{\mov(3,1){\lin(-2.1,2.1)\lin(2.1,2.1)\dashlin(0,-2)}
\mov(8,1.5){\lin(1,0)\arrow(4,0)}
\mov(15,1.5){$\hat Z$}
\mov(18,0){\lin(-3.1,3.1)\lin(3.1,3.1)\dashlin(0,-1)
\mov(-2.75,2.5){\dashlin(5,0)}
\mov(-2.5,1){$\alpha$}\mov(+1,1){$\alpha$}
\mov(-1.2,2.8){$\beta$}}}
\caption{The conformal triangle.\hspace*{5cm} {}}
\label{triangle}
\end{figure}
\begin{enumerate}
\item replace in the base diagram
$ G $ the bare propagators by dressed ones
and the pointlike triple
vertex by the conformal triangle~(see Fig.~{\ref{triangle}}):
\begin{eqnarray}
\label{AB}
&&\hat G_{\phi}(x) \equiv
\frac{\hat A}{x^{2\Delta_{\phi}}},
\ \ \ \ \ \  \hat G_{\psi}(x) \equiv
\frac{\hat B}{x^{2\Delta_{\psi}}},\\
&&\gamma_{R}(x,y;z) \equiv
\hat Z (z-x)^{-2\alpha}(z-y)^{-2\alpha}(x-y)^{-2\beta}.
\end{eqnarray}
Here
$ \al=\mu-1-\gamma_{\psi}/2 $,
$ \bet=2-\gamma_{\phi}+\gamma_{\psi}/2 $
and
$ \Delta_{\phi}, \Delta_{\psi} $ are the full scaling dimensions of the
fields
$ \phi, \psi $, respectively.
$ \Delta_{\phi}=\mu-1+\gamma_{\phi}, $
$ \Delta_{\psi}=2+\gamma_{\psi}. $

\item introduce a regulator
$ \Delta $ in anyone line (It is needed because after
the above changes one obtains
a superficially divergent diagram).
\item
evaluate the $ \Delta $ pole residue in the
resulting diagram and pick out the \NN order term -- $G^{(2)}$.
Then the required contribution is given by
$ -2G^{(2)} $.
\end{enumerate}
The  explicit
expressions for the amplitudes  $\hat A$ , $\hat B$ ,
$\hat Z$ are given in Appendix A.

Following these prescriptions one need calculate only one~--"conformal"
diagram~(see Fig.~\ref{CONF})~--~ instead of the six \NN order diagrams
shown on Fig.~\ref{FN2}~(c-h). It should be noted also, that the
resulting diagram, unlike the original ones, has  superficial divergence
only, that, in its turn, gives the definite advantages at the
calculations. We end this section with remark, that the above procedure
being useful even if the condition
$ 2n_{\psi}=n_{\phi}=n_{V} $  is not fulfilled.
Indeed, the only changes which will occur are that the last sum in the
Eq.~(\ref{sum2}) will not be zero any longer and will contribute to the
anomalous dimension on the equal foot with the first ones.

\section{The  anomalous dimensions of the
$ (\otimes \vec \Phi)^{s}$ and
$ (\vec \Phi\otimes(\otimes \vec\df)^{n} \vec \Phi)$
operators in
$ 1/N^{2} $ order}
\label{ff}

In this section we apply the technique developed above to calculate
the anomalous dimensions of the  operators
\begin{eqnarray}
F^{s}_{1}&=&(\otimes \vec \Phi)^{s}\equiv
\mbox{Sym}\phi^{A_{1}}\ldots\phi^{A_{s}}\  - \ \mbox{traces},\\
F^{n}_{2}&=&(\vec \Phi\otimes(\otimes \vec\df)^{n} \vec \Phi)\equiv
\mbox{Sym} \phi^{A}\df_{i_{1}}\ldots\df_{i_{n}} \phi^{B}\ - \
\mbox{traces}
\end{eqnarray}
in the \NN\ order.  Here, the symbol
$ \mbox{Sym} $  implies the symmetrization over the internal indeces as
well as the spatial ones.

\begin{figure}
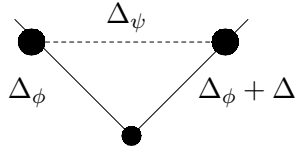

\Lengthunit=0.5cm

\GRAPH(hsize=10){
\mov(15,0){\mov(-2.4,2.5){\Circle**(0.7)}\mov(2.75,2.5){\Circle**(0.7)}
\Circle**(0.5)\lin(-3.1,3.1)\lin(3.1,3.1)
\mov(-2.75,2.5){\dashlin(5,0)}
\mov(-3.8,1){$\Delta_{\phi}$}
\mov(1,1){$\Delta_{\phi}+\Delta$}
\mov(-1.7,3){$\Delta_{\psi}$}
}}
\caption{The "conformal" diagram. The black circles
denote the conformal triangle.}
\label{CONF}
\end{figure}

\begin{figure}
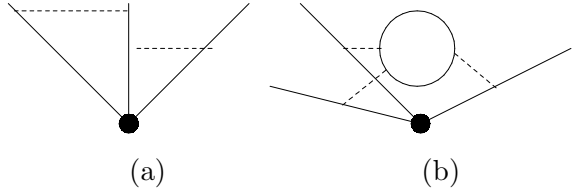

\Lengthunit=0.5cm

\GRAPH(hsize=10){
\mov(10,0){\Circle**(0.5)\lin(-3.2,3.2)\lin(3.2,3.2)\lin(0,3.2)
\mov(-3.3,3){\dashlin(3,0)}
\mov(-0.3,2){\dashlin(2,0)}\mov(-0.5,-1.5){(a)}}
\mov(17.5,0){\Circle**(0.5)\lin(-3.2,3.2)\lin(-4,1)\lin(4,2)
\mov(-2.3,2){\dashlin(1,0)
\mov(-0.3,0){\mov(2,0){\Circle(2)}}}
\mov(1.5,0.95){\dashlin(-1.1,0.92)}
\mov(-3.1,0.5){
\dashlin(1.15,0.92)}
\mov(-1,-1.5){(b)}}}
\caption{The
$ 1/N^{2} $ order diagrams with insertion of operator
$F_{1}$.}
\label{FN3}
\end{figure}

The operators
$ F_{1} $   are multiplicatively renormalized, whereas
$ F_{2} $ does not. However, the operators admixing to
$ F_{2} $ being the total derivatives, they are irrelevant for our
purposes.  The diagrams to be evaluated for the determination of the
anomalous dimensions of operators
$ F_{1}, F_{2} $ are those shown on Fig.~\ref{FN2} and Fig.~\ref{FN3},
the latter being obviously relevant only for the
$ F_{1} $ operators. Because the diagrams
$ c-h $ on Fig.~\ref{FN2} can be replaced by one "conformal" diagram
(Fig~\ref{CONF}) in accordance with the rules formulated in the previous
section, one has to calculate $5$ diagrams for the
$ F_{1} $ operators and
$ 3 $  ones for the
$ F_{2} $ operators.

We do not dwell upon the technique of the calculating of the diagrams,
since it is discussed in very details in the
papers~\cite{V1,V2,V3,kaz,kazkot} and give the final results for the
anomalous dimensions only. (For completeness, we give the explicit
expressions for the pole terms of each diagram in Appendix~\ref{Ab}
and the some useful formulae are collected in Appendix~\ref{Ac}.)

The anomalous dimensions
$ \gamma_{(s)} $ and
$ \gamma_{(n)} $ of the operators
$ F_{1}^{s} $ and
$ F_{2}^{n} $ it is convenient to represent as
\begin{equation}
\gamma_{(s)}=u_{(s)}+s\gamma_{\phi}\ \ \ \ \ \ \
\gamma_{(n)}=u_{(n)}+2\gamma_{\phi},
\end{equation}
where
$ \gamma_{\phi} $ is the anomalous dimension of the field
$ \phi $ known up to
$ 1/N^{3} $ order~\cite{V3}.
Then representing
$ u $ in the following form
$ u=\sum_{k=1}^{\infty}u^{(k)}/N^{k} $ we have for the two first terms:
\begin{eqnarray}
u^{1}_{(s)}&=&-\frac{s(s-1)\mu}{2(\mu-2)}\eta_{1},\\
u^{2}_{(s)}&=&-{\eta_1^2} \frac{s(s-1)\mu}{4(\mu-1)(\mu-2)^2}\Bigl \{
2(s-2)\mu(\psi^{\prime}(1)-\psi^{\prime}(\mu))
      \nonumber\\
&&+[\mu(2\mu-3)+2(\mu-1)(2\mu^2-3\mu+2)R]  \Bigr\},
\end{eqnarray}
\begin{eqnarray}
u^{1}_{(n)}&=&-\frac{\mu(\mu-1)}{(\mu-1+n)(\mu-2+n)}{\eta_1},\\
u^{2}_{(n)}&=&-\eta_{1}^{2}\frac{\mu(\mu-1)}{(\mu-1+n)(\mu-2+n)}
\Bigl \{\frac{(2\mu^2-3\mu+2)}{\mu-2}R -
\frac{2 (\mu-1)(2\mu-1)}{\mu-2}S(n,\mu) \nonumber\\
&&+\frac12\mu(\mu-1)R(n,\mu)+
\frac{1}{\mu-2}\left ( \frac{n}{\mu-2+n}+
\frac{\mu(\mu^2-5\mu+5)}{(\mu-1)(\mu-2)}\right) \nonumber\\
&&-\frac{\mu(\mu-1)}{2(\mu-1+n)(\mu-2+n)}\left
(1-\frac{1}{(\mu-1+n)}-\frac{1}{(\mu-2+n)}\right)  \Bigr \},
\end{eqnarray}
where
\begin{eqnarray}
&& \eta_{1}=
4(2-\mu){\Gamma(2\mu-2)}/{\Gamma^{2}(\mu-1)\Gamma(2-\mu)}
{\Gamma(\mu+1)}, \label{eta1}\\
&&R= \psi(1)+\psi(\mu-1)-\psi(2-\mu)-\psi(2\mu-2),
\label{RR}\\
&&S(n,\mu) = \psi(\mu-1+n)-\psi(\mu-1),\\
&&R(n,\mu)=
\int_0^1\int_0^1 {\rm d}\alpha{\rm d}\beta
\alpha^{\mu-3}\beta^{\mu-3}(1-\alpha-\beta)^n.
\end{eqnarray}

The critical exponents of the operators $F_{1},F_{2}$ are known with the
high accuracy in the
$ 4-\eps $ as well as in the
$ 2+\eps $ expansions.  Expanding our result for
$ \gamma_{(s)} $ in the
$ \eps $  series near
$ d=2 $ we have obtained the full agreement with the four~--~loop result
of Wegner~\cite{W4loop}. The same exponents are known
up to
$ \eps^{3 }$ order in the
$ 4-\eps $ expansion~\cite{WalZia}.  In this case  we find out the
discrepancy of our result with those of Wallace and Zia~\cite{WalZia}
in term
$ \sim\eps^{3}s(s-1)/N^{2} $ (in our answer this term enters with
the coefficient
$(-10) $, against the coefficient $ (-8) $ in the paper~\cite{WalZia}).

Further, using the results of the paper~\cite{DJM}, where the
four~--~loop  anomalous dimensions of
$ \phi(\otimes\vec \df)^{n}\phi $ operators had been calculated in the
scalar
$ \phi^{4} $ model in the
$ 4-\eps $  expansion, we reconstructed the answer for the
$ O(N) $ vector model by restoring the
$ O(N) $ coefficients for all diagrams. Comparing the expressions for
the critical exponents of
$ F_{2} $ operators derived in the both
($ 1/N $ and $ 4-\eps $)
approaches we found that they are in the full agreement up to
$ \eps^{4}/N^{2} $  terms.

\section{Conclusion}

In the present paper we have developed the simple method for the
calculating of the anomalous dimensions (up to \NN\ order) in the \N\
expansion. In the comparison with the other well known methods of the
conformal bootstrap and the selfconsistency equations this one has the
advantage of being not restricted to the special class of the operators.
In this approach the formulae relating the anomalous dimensions with the
divergencies of the corresponding Feynman diagrams are very simple and
in the one to one correspondence with their counterparts in the
dimensional regularization scheme. These features together with the
uniqueness of the triple vertex of the nonlinear sigma model make the
\NN\ order calculations of the rather straightforward task.
In some cases
(see for example the sec.~\ref{ff}) the calculations are not much more
complicated than the two~--~loop ones in the
$ O(N) $ model in the
$ 4-\eps $ expansion.  However, in a general situation, beyond
\N\ order,
the number of the diagrams to be evaluated is usually large enough.  We
have shown that this problem is partially removed after going on to the
diagrams with the dressed propagators and vertices.
The number of the diagrams is considerably reduced by this trick
and only slightly exceeds the number  of
those in the conformal bootstrap method,
which is considered to be the most economic in this sense.

We conclude with remark, that though so far we consider the sigma model
only, it is evident that the same technique can be apply to the many
other models admitting \N expansion~---~
$ CP(N) $, Gross~--~Neveu and Thirring models, etc.
\vskip 1cm

\centerline{\Large \bf Acknowledgments}
\vskip 0.5cm

It is a pleasure to thank
Prof.~A.~N.~Vasil'ev  for the valuable
discussions and Dr.~J.~Gracey for critical remarks.
The work by S.D. was supported by NORDITA as a part of
the Baltic Fellowship program funded by the Nordic Council of Ministers.
This work was also supported in part by  the
Russian Fond of Basic Research (Grant 97--01--01152)
and by INTAS Grant
93--2492--ext and carried out under the research program of the
International Center for Fundamental Physics in Moscow.

\newpage

\appendix
{\Large \bf Appendices\hfil}

\sect{The renormalized Green functions in $1/N$ order}
\label{Aa}
In this appendix we give the explicit expressions for the amplitudes
$ \wh A $, $ \wh B $, $ \wh Z $ introduced into consideration in
the Sec.~\ref{confors}.
Obviously, these quantities depend on the our choice of the function
$ C(\Delta) $~(see Eqs.~(\ref{propP}),(\ref{propX})).
However, it is easy
to understand, that in the first \N\ order this dependence
is almost trivial
and can be taking into account by the redefinition of the parameter
$ M $. We specify
$ C(\Delta) $ by the condition for the propagator of
$ \psi $ field in coordinate space~(Eq.(\ref{propX})) to have
the amplitude independent from
$ \Delta $, except from the factor
$ M^{2\Delta} $.
Then one has for the amplitude
$ A,B $ of the bare propagators in the coordinate space:
\begin{equation}
A=\frac{H(1)}{4\pi^\mu},\ \ \ \ \
B=-M^{2\Delta}\frac{32H(2-\mu)}{H(2)H^{2}(1)}.
\label{abb1}
\end{equation}

Evaluating the \N order diagrams for
$ 1PI $  Green functions
$\Gamma_{\phi\phi}$ , $\Gamma_{\psi\psi}$
$\Gamma_{\psi\phi\phi}$ shown on Fig.~\ref{apd}
we obtain for the amplitudes $ \wh A $,$ \wh B $,$ \wh Z $:
\begin{figure}
\Lengthunit=0.7cm
\GRAPH(hsize=10){
\mov(3,0){
$\Gamma_{\phi\phi} = - G_{\phi}^{-1} \ +$
\mov(0.2,0.1){\lin(2,0)\mov(0.1,0){\dasharcto(1.8,0)[+1.5]}}
\mov(2.5,0){
$;\ \Gamma_{\psi\phi\phi} = 1 \ +$
\mov(0.1,0.1){\lin(1,0.5)\lin(1,-0.5)\mov(0.9,0.4){\dashlin(0,-0.8)}}
\mov(1.1,0){$+\ $}
\mov(1.6,0.1){
\lin(1,0.5)\lin(1,-0.5)\mov(1,0.5){\lin(0,-1)}
\mov(0.8,0.5){\dashlin(1,0)}\mov(0.8,-0.5){\dashlin(1,0)}
\mov(1.6,0.5){\lin(0,-1)\lin(0.1,0.05)}
\mov(1.6,-0.5){\lin(0.1,-0.05)}}}}}
\vspace{0.3cm}
\GRAPH(hsize=10){
\mov(3,0)
{
$\Gamma_{\psi\psi} = - G_{\psi}^{-1} \ +\ \frac{1}{2}$
\mov(0,0.1)
{
\mov(0.05,0){\lin(0.15,0)}
\mov(0.5,0){\Circle(1)\mov(0,0.5){\dashlin(0,-1)}}
\mov(1,0){\lin(0.15,0)}
}
\mov(1.5,0)
{
$+\ $\mov(0.05,0.1){\lin(0.15,0)}
\mov(0.5,0.1){\Circle(1)\mov(-0.45,0.2){\dashlin(0.9,0)}}
\mov(0.8,0.1){\lin(0.15,0)}
}
\mov(3.6,0){$+\ \frac{1}{2}$}
\mov(4.6,0.1){
\mov(0.05,0){\lin(0.15,0)}
\lin(1,0.5)\lin(1,-0.5)\mov(1,0.5){\lin(0,-1)}
\mov(0.8,0.5){\dashlin(1,0)}\mov(0.8,-0.5){\dashlin(1,0)}
\mov(1.6,0.5){\lin(0,-1)\lin(1,-0.5)}\mov(1.6,-0.5){\lin(1,0.5)}
\mov(2.4,0){\lin(0.15,0)}
}}}
\caption{The
$ 1/N $ order diagrams for the Green functions
$\Gamma_{\phi\phi}$ , $\Gamma_{\psi\psi}$ and
$\Gamma_{\psi\phi\phi}$.}
\label{apd}
\end{figure}
\begin{eqnarray}
\wh A &=& AM^{-2\gamma_{\phi}} \cdot
\left[1 - \frac{\eta_{1}}{2N}\frac{\mu^2+\mu-1}{\mu(\mu-1)}
\right], \\
\wh B &=& BM^{-2\gamma_{\psi}} \left[
1 + \frac{\eta_{1}}{N} \left(
\frac{2\mu^2-3\mu+2}{\mu-2}\cdot R -
\frac{4\mu^5-19\mu^4+25\mu^3-\mu^2-14\mu+4}{\mu(\mu-1)(\mu-2)^2}
\right)
\right],\\
\wh Z &=& -M^{-2\chi}\frac{\chi}{2}\frac{H^2(1) H(\mu-2)
\Gamma(\mu)}{\pi^{2\mu}} \cdot \left[ 1 + \frac{\eta_{1}}{N}
\frac{(\mu-3)(6\mu^2-9\mu+2)}{(\mu-2)^2}\right].
\end{eqnarray}
Here
$ \eta_{1},R $  are given by Eq.~(\ref{eta1}),(\ref{RR}) and
$ \gamma_{\phi}, \gamma_{\psi} $ are the known anomalous dimensions of
the fields
$ \phi,\psi $
\begin{eqnarray}
\gamma_{\phi}&=&\frac{\eta_{1}}{2N}+O(1/N^{2}), \nonumber \\
\gamma_{\psi}&=&-\frac{2\eta_{1}}{N}\frac{(\mu-1)(2\mu-1)}{2-\mu}+
O(1/N^{2}).
\end{eqnarray}
The index
$ \chi $  is defined as
$ \chi=-(\gamma_{\psi}+2\gamma_{\phi}). $

\sect{Values for the $1/N^{2}$ order integrals}
\label{Ab}

In this appendix we give the explicit expressions for the contributions
to the anomalous dimensions
$ \gamma_{(s)}, \gamma_{(n)} $ of the corresponding Feynman integrals.
We will  use the labels
$ A,B,C,D,E $  for the diagrams on the Fig.~\ref{FN2}~a,b,
Fig.~\ref{CONF} and Fig.~\ref{FN3}~a,b, resp., as well as for the
values of the latters. In the case of the
$ F_{1}^{(s)} $ operators we have obtained for the simple
$ \Delta $ pole residue of the corresponding integrals
in the units
$ \eta_{1}^{2}/N^{2} $:
\begin{eqnarray}
A&=&
- \frac{s(s-1)}{2}
\frac{\mu^2(\mu^2-5\mu+5)}{8(\mu-1)(\mu-2)^3}, \\
&&\nonumber \\
B&=&
\frac{s(s-1)}{2}
\frac{\mu^2(\mu-1)}{8(\mu-2)^3},   \\
&&\nonumber \\
C&=&
s(s-1)\frac{\mu}{(2\mu-2)^2}\left[
\frac{\mu(\mu^2-5\mu+5)}{(\mu-1)(\mu-2)} + (2\mu^2-3\mu+2)R
\right],\\
&&\nonumber \\
D&=&
{s(s-1)(s-2)}
\frac{\mu^2}{8(\mu-1)(\mu-2)^2},\\
&&\nonumber \\
E&=&
\frac{s(s-1)(s-2)}{6}\frac{3\mu^2(\mu-1)}{2(\mu-2)^2}C.
\end{eqnarray}
For diagrams with the insertion of the
$ F_{2}^{(n)} $  operators we derived
\begin{eqnarray}
A&=&
- \frac{\mu^2(\mu-1)^2}{8(\mu-1+n)^2(\mu-2+n)^2}
\left(1-\frac{1}{\mu-1+n}-\frac{1}{\mu-2+n}\right),\\
&&\nonumber \\
B&=&
\frac{\mu^2(\mu-1)^2 }{8(\mu-1+n)(\mu-2+n)}R(n;\mu),\\
&&\nonumber \\
C&=&
\frac{\mu(\mu-1)}{2(\mu-1+n)(\mu-2+n)}
\Bigl \{\frac{(2\mu^2-3\mu+2)}{\mu-2}R -\nonumber\\
&&\nonumber \\
&&-\frac{2 (\mu-1)(2\mu-1)}{\mu-2}S(n,\mu)+
\frac{1}{\mu-2}\left ( \frac{n}{\mu-2+n}+
\frac{\mu(\mu^2-5\mu+5)}{(\mu-1)(\mu-2)}\right)
\Bigr \}. \nonumber\\
\end{eqnarray}
Each of these quantities enters into expression for the
anomalous dimension with the coefficient ($ -2n_{\psi} $),
if we accept
$ n_{\psi}=1 $  for diagram
$ C $.

\section{Chain rule and uniqueness relation}
\label{Ac}
Here we collect some rules~(see Refs.~\cite{V2,V3,kaz})
which form the basis for
the calculations of diagrams in the massless theories.
As usual, we will denote the propagator
$ 1/|x-y|^{2\al} $  by the line with the index
$ \al $.

The Fourier-transform of
$ |x|^{-2\al} $ reads
\begin{equation}
 \int d^{2\mu}x\frac{\mbox{e}^{\imath px}}{x^{2\al}}=
\pi^{\mu}\frac{2^{2\al^{\prime}}H(\al)}{p^{2\al^{\prime}}}.
\end{equation}
We remind that
 $\mu \equiv D/2$,  $H(z) \equiv \Gamma(\mu-z)/\Gamma(z)$
and $z^{\prime} \equiv \mu - z$.

Further, we consider the chain of propagators, which is illustrated on
Fig.~\ref{APchain}. The point where the ends of propagators are joined
is the vertex of integration.
The result of this integration is a new line with index
$ \al+\bet-\mu $ and amplitude proportional to
$ \pi^{\mu}H(\al,\bet,2\mu-\al-\bet) $, where
$ H(\al_1,\al_2,\al_3)=\prod^{3}_{i=1}H(\al_i) $.
\begin{figure}
\Lengthunit=0.7cm
\GRAPH(hsize=10){\mov(5.,0.){
\lin(1.7,0)
\mov(0.5,0.1){$\alpha$}
\mov(1.4,0){\lin(1.5,0)\Circle**(0.1)
\mov(0.5,0.1){$\beta$}}
\mov(3.1,-0.1){$= \pi^{\mu} H(\alpha,\beta,2\mu-\al-\bet)$}
\mov(9,0){\lin(3,0)\mov(0.5,0.1){$\al+\bet-\mu$}}}}
\vskip 1cm
\GRAPH(hsize=10){\mov(7.,0.){
\mov(0.2,0){\Circle**(0.1)\lin(0,1.4)\mov(0.1,0.5){$\gamma$}}
\lin(-1,-1)\mov(-0.9,-0.5){$\alpha$}\lin(1,-1)\mov(0.6,-0.5){$\beta$}
\mov(1.2,0){$ = \pi^{\mu} H(\alpha,\beta,\gamma)$}
\mov(6.5,1.4){\lin(-1,-2.4)\lin(1,-2.4)\mov(-1,-2.4){\lin(2,0)}}
\mov(7,0){$\alpha^{\prime}$}\mov(5.2,0){$\beta^{\prime}$}
\mov(6,-1.5){$\gamma^{\prime}$}
\mov(1,-2.5)
{$ \al+\bet+\gamma=2\mu $}}}
\caption{Chain rule and uniqueness relation}
\label{APchain}
\end{figure}

The other useful  relation is the so-called uniqueness relation.
In the case if the sum of the exponents of the lines meeting at the
vertex
is equal to the dimension of space-time it can be related to a
triangle
graph~(see Fig.~\ref{APchain}).

The corresponding generalization of these rules needed for the
evaluation of the the diagrams with the insertion of the tensor
operators,
are given in Refs.~\cite{kazkot}. We give here only that which is
relevant  for the treatment of the traceless and symmetric operators,
like
 operator $ F_{2} $ in our case.
It is convenient to swamp the
 the spatial indices $\nu_{1},...,\nu_{n}$ of a operator
by contracting with a constant vector $u^i$, such that
$ u^{2}=0 $, and consider scalar operator
$ T_{n}(x,u)=T_{n}^{\nu_1\cdots\nu_{n}}(x)
u_{\nu_1}\ldots u_{\nu_{n}} $.
In the course of calculations one should often to integrate the chains
of the propagators like
$ (ux)^{k}/|x|^{2\al} $. Then, the corresponding rule for the
integration
of the chains   reads
\begin{equation}
\int {{\rm d}^{2\mu} x}
\frac{(x,u)^{n} (z-x,u)^{m}}{x^{2\alpha}(x-z)^{2\beta}} =\pi^{mu}
\frac{H_n(\alpha) H_m(\beta)}{H_{n+m}(\alpha+\beta-\mu)}
\frac{(u, z)^{m+n}}{z^{2(\alpha+\beta-\mu)}},
\end{equation}
where
$ H_n(z) \equiv {\Gamma(\mu+n-z)}/{\Gamma(z)} $.

\end{document}